\documentclass{article}

\usepackage[T1]{fontenc}
\usepackage[utf8]{inputenc}
\usepackage{cite}
\usepackage[preprint]{neurips_2025}
\usepackage{url}
\usepackage{listings}
\usepackage{amsmath}
\usepackage{microtype}
\usepackage{hyperref}
\usepackage{tikz}
\usetikzlibrary{arrows.meta,positioning,decorations.pathreplacing,calc}

\begin{document}

\title{Bootstrapping Coding Agents: The Specification Is the Program}

\author{Martin~Monperrus%
\thanks{M.~Monperrus is with KTH Royal Institute of Technology, Stockholm, Sweden.
E-mail: monperrus@kth.se}%
}


\maketitle


\begin{abstract}
A coding agent can bootstrap itself. Starting from a 926-word specification
and a first implementation produced by an existing agent (Claude Code),
a newly generated agent re-implements the same specification correctly from
scratch. This reproduces, in the domain of AI coding agents, the classical
bootstrap sequence known from compiler construction, and instantiates the
meta-circular property known from Lisp. The result carries a practical
implication: the specification, not the implementation, is the stable
artifact of record. Improving an agent means improving its specification;
the implementation is, in principle, regenerable at any time.
\end{abstract}

\section{Introduction}

Coding agents are programs that accept natural-language task descriptions and
produce or modify source code. Teams now deploy them to write tests, perform refactoring, and
implement features from natural language task descriptions and requirements~\cite{lopopolo2026,jimenez2024}.

Compiler writers discovered decades ago that a new language implementation
passes a meaningful milestone when it can compile itself. This property,
called \emph{self-hosting}, is not merely a curiosity: it validates that the
implementation is expressive enough to describe its own behavior. A
self-hosting compiler is also a fixed point of the compilation process: it
reproduces itself under its own translation.

The same milestone has now been reached for AI coding agents. Given only a
natural-language specification, a coding agent can implement itself. This article
describes the experiment, draws the analogy to classical results in computer science, and examines the implications for software engineering
practice.

\section{The Bootstrapping Experiment}

The experiment consists of three steps, each building on the previous one. All
artifacts are publicly available in the \texttt{meta-circular} repository~\cite{metacircular}.

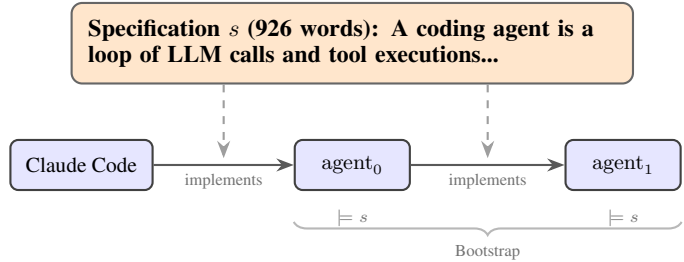
\begin{figure}[t]
\centering
\begin{tikzpicture}[scale=0.9, transform shape,
  >=Stealth, thick,
  agbox/.style={rectangle, rounded corners=3pt,
    draw=black!75, fill=blue!10,
    minimum width=1.7cm, minimum height=0.74cm, font=\small},
  ccbox/.style={rectangle, rounded corners=3pt,
    draw=black!75, fill=blue!10,
    minimum width=2.1cm, minimum height=0.74cm, font=\small},
  specbox/.style={rectangle, rounded corners=4pt,
    draw=black!60, fill=orange!20, thick,
    text width=7.8cm, align=left, inner sep=7pt},
  arr/.style={->, thick, black!65},
  darr/.style={->, thick, dashed, black!38},
  lbl/.style={font=\scriptsize, text=black!55},
]
  \node[specbox] (spec) at (4.0, 1.8) {%
    \textbf{Specification $s$ (926 words): A coding agent is a loop of LLM calls and tool executions...}
  };

  \node[ccbox] (cc) at (0,   0) {Claude Code};
  \node[agbox] (a0) at (4.0, 0) {$\mathrm{agent}_0$};
  \node[agbox] (a1) at (8.0, 0) {$\mathrm{agent}_1$};

  \draw[arr] (cc) -- node[below, lbl] {implements} (a0);
  \draw[arr] (a0) -- node[below, lbl] {implements} (a1);

  \draw[darr] ([xshift=-1.9cm]spec.south) -- (2.1, 0.10);
  \draw[darr] ([xshift= 2.0cm]spec.south) -- (6.0, 0.10);

  \node[lbl, below=4pt of a0] {$\models s$};
  \node[lbl, below=4pt of a1] {$\models s$};

  \draw[decorate, decoration={brace, amplitude=5pt, mirror}, black!28]
    ($(a0.south west)+(0,-13pt)$) -- ($(a1.south east)+(0,-13pt)$)
    node[midway, below=6pt, lbl] {Bootstrap};
\end{tikzpicture}
\caption{The coding-agent bootstrap sequence. Claude Code implements
$\mathrm{agent}_0$ from the specification; $\mathrm{agent}_0$ then
bootstraps $\mathrm{agent}_1$ from the same specification. Dashed arrows
show where the specification is consumed. Both implementations
satisfy~$s$, denoted~$\models s$.}
\label{fig:bootstrap}
\end{figure}

\subsection{Step 1: Specification}

A large language model (LLM) was prompted to write a specification for a coding
agent: a program that receives a task description in natural language and
produces or modifies source code to perform the task. The resulting document
defines the agent's interface, its expected behavior, and the constraints it
must respect. The specification is 926 words long and is available at the
repository listed in the references~\cite{metacircular}.
It covers three areas: the agent's \emph{interface} (command-line arguments,
environment variables, and API interaction), its \emph{behavioral constraints}
(how it handles multi-turn conversations, tool use, and error conditions), and
the \emph{tool loop} (the cycle of receiving a task, calling tools, observing
results, and producing output). The document is intentionally compact: at 926
words, it can be read and audited by a single engineer in under five minutes.

\subsection{Step 2: First Implementation}

Claude Code with Sonnet 4.6~\cite{claudecode}, an existing coding agent, was given the
specification and asked to implement it. The prompt was minimal:

\begin{lstlisting}[language={}]
implement the spec in a single python file
\end{lstlisting}

The result is a working Python program (\texttt{agent.py}) that satisfies the
specification. No manual edits were required. The program accepts a task
description as a command-line argument and uses a language model API to
produce or modify source code:

\begin{lstlisting}[language={}]
$ python agent.py
usage: agent.py [-h] [--model MODEL]
  [--base-url BASE_URL] [--api-key API_KEY]
  [--max-turns MAX_TURNS] [--cwd CWD]
  [task]
\end{lstlisting}

\subsection{Step 3: Self-Implementation}

The newly generated agent was given the same specification and asked to
implement it again, using the same prompt as Step~2. It succeeded. The
agent implemented itself:

\begin{lstlisting}[language={}]
$ python agent.py \
    "implement the spec in a single python file"
\end{lstlisting}

The output is a new \texttt{agent.py} that satisfies the same specification.
Both the first-generation and second-generation programs were verified manually
against the specification (Fig.~\ref{fig:bootstrap}).

\section{Meta-Circularity}

The term \emph{meta-circular} comes from the Lisp tradition. In 1960, McCarthy
presented a Lisp interpreter written in Lisp itself~\cite{mccarthy1960}. The
interpreter is called meta-circular because the language in which it is written
is the same language it interprets. The implementation and the object of
interpretation are drawn from the same formalism.

The first practical self-hosting Lisp compiler was implemented by Hart and
Levin at MIT in 1962~\cite{hartlevin1962}, two years after McCarthy's
theoretical definition. Their work established the bootstrap sequence that
compiler writers would follow for decades: define the language formally,
implement it in another host, then use that implementation to compile itself.
More recently, Courant et al.\ reproduced the bootstrap for OCaml using the
Camlboot project, demonstrating that a minimal, auditable compiler can
found its full industrial counterpart~\cite{courant2022}. This creates a
clear lineage between the meta-circularity definition (McCarthy 1960), to the coding-agent bootstrap (2026).

The coding-agent experiment follows the same pattern. The agent generated in
Step~2 is, in this terminology, a meta-circular agent: it is an implementation
of ``coding-agent behavior'' produced by a system that itself exhibits
``coding-agent behavior.'' When the Step-2 agent reproduces itself in Step~3,
it demonstrates meta-circularity in the operational sense. The agent is a fixed
point of its own implementation procedure.

To recap, let $F$ be the function that maps a specification and a coding agent to an
implementation satisfying that specification. Let $s$ be the coding-agent
specification. Then:
\begin{align}
  \mathrm{agent}_0 &= F(s,\ \text{Claude Code}) \label{eq:a0}\\
  \mathrm{agent}_1 &= F(s,\ \mathrm{agent}_0) \label{eq:a1}
\end{align}

This is illustrated in \autoref{fig:bootstrap}. This is the coding-agent analogue of a
self-hosting compiler: the first generation is sufficient for the second
generation to be indistinguishable from it under the specification. The
specification, not the bootstrap tool, is the stable artifact.

\begin{figure}[t]
\centering
\begin{tikzpicture}[scale=0.9, transform shape,
  >=Stealth, thick,
  cbox/.style={rectangle, rounded corners=3pt,
    draw=black!70, fill=green!10,
    minimum width=1.35cm, minimum height=0.66cm, font=\small},
  abox/.style={rectangle, rounded corners=3pt,
    draw=black!70, fill=blue!10,
    minimum width=1.35cm, minimum height=0.66cm, font=\small},
  arr/.style={->, thick, black!65},
  lbl/.style={font=\scriptsize, text=black!55},
  hdr/.style={font=\small\bfseries, text=black!70},
]
  \node[hdr]  at (-0.5, 1.15) {Compiler};
  \node[cbox] (c0) at (1.5,  1.15) {$C_0$};
  \node[cbox] (c1) at (5.5,  1.15) {$C_1$};
  \node[lbl]  at   (7.3,  1.15) {$C_1 \equiv C_0$};
  \draw[arr] (c0) -- node[above, lbl] {self-compile} (c1);

  \draw[black!18, thin] (-1.1, 0.57) -- (8.4, 0.57);

  \node[hdr]  at (-0.5, 0.0) {Agent};
  \node[abox] (a0) at (1.5,  0.0) {$\mathrm{agent}_0$};
  \node[abox] (a1) at (5.5,  0.0) {$\mathrm{agent}_1$};
  \node[lbl]  at   (7.3,  0.0) {$\mathrm{agent}_1 \equiv \mathrm{agent}_0$};
  \draw[arr] (a0) -- node[above, lbl] {self-impl.} (a1);
\end{tikzpicture}
\caption{Structural analogy between the classical compiler bootstrap (top)
and the coding-agent bootstrap (bottom). The first self-generated artifact is equivalent to the
first-generation one.}
\label{fig:analogy}
\end{figure}
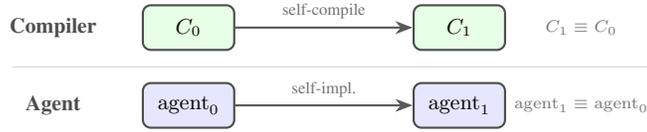

\section{Implications}

\subsection{Specification as the Focus of Correctness}

If a specification is sufficient to generate an agent, and the agent is
sufficient to regenerate itself from the same specification, then the
specification is the true focus of correctness. Improving the agent means
improving the specification. The implementation is, in principle, disposable
and regenerable at any time.

This reframes code review. Rather than auditing every line of an
auto-generated implementation, reviewers should audit the specification from
which the implementation is derived~\cite{osmani2026}. Errors in the
specification propagate to every generation of the implementation or to every generation in different programming languages.

An industrial-scale instance of this pattern is described by Lopopolo et
al.~\cite{lopopolo2026}: a team of three to seven engineers built a
million-line codebase using Codex over five months, with zero manually written
lines of code. The team treats their structured \texttt{docs/} directory as the
reference system, not the code itself. The codebase is regenerable; the
specification is the stable artifact. This mirrors, at industrial scale, the
implication derived from the bootstrap experiment.

This result aligns with the emerging practice of Spec-Driven
Development~\cite{piskala2026}, which treats the specification as the primary
engineering artifact and generated code as a build product. The practical
consequences are concrete. In version control, the specification belongs
on the main branch and is subject to the full review workflow; generated
implementation files may be treated as build artifacts, reconstructed on demand
rather than tracked commit-by-commit. In team workflows, specification
review replaces code review as the primary quality gate. An engineer who can
evaluate whether a specification correctly and completely describes the intended
behavior provides more durable value than one who audits lines of generated
code that may be regenerated wholesale by the next model update.

\subsection{Good Specifications for AI Coding}

The bootstrap result raises a practical question: what properties must a
specification have to serve as the stable artifact of record?

\emph{Size and auditability.} The 926-word specification used here is small
enough to read in a single sitting. A useful budget: keep the specification
under 1,500 words, or under 15 minutes of reading time. If a section requires
cross-referencing another section to be understood, split or restructure.

\emph{Behavioral completeness.} The specification must describe every
capability the agent exercises. A practical checklist: for each tool call, API
interaction, and error condition in the generated code, ask ``is this behavior
described in the specification?'' Any ``no'' is a gap. Write the specification
first and treat undescribed behavior as intentionally undefined.

\emph{Convergence across implementations.} A good specification is unambiguous
enough that two independent implementations converge on the same observable
behavior. Heuristic: give the specification to two different engineers
(or two independent LLMs ~\cite{ron2024}) and check whether the results agree on
externally visible behavior such as command-line interface and output format.
This is the N-version programming principle applied to specification
validation~\cite{avizienis1985}: divergence between independently
produced implementations reveals ambiguity that must be resolved in the
specification.

\emph{Abstraction level.} A specification should describe \emph{what} the
agent does, not \emph{how}. If a sentence names a data structure, a library,
or a language feature, it is probably too low-level. Sentences that name only
inputs, outputs, and conditions are at the right level and will remain stable
across model generations.

\subsection{Trusting trust \& AI Coding}

Thompson's 1984 Turing Award lecture showed that a trojan horse in a compiler
binary can reproduce itself through every subsequent compilation without
appearing in any source file~\cite{thompson1984}. The coding-agent bootstrap
introduces an analogous structure. A compromise in the model weights, the API
endpoint, or the execution environment can propagate silently through every
generated implementation. Like a compiler binary, model weights are opaque and are updated out-of-band by a third party.

The practical response is to treat each layer as a versioned, auditable
dependency. Pin the model version used for generation, run generation in a controlled
CI environment whose executions are fully logged, record the full trajectories. The specification remains the
only artifact that a human engineer can read and audit directly; everything
downstream of it is a build product that should be reproducible from a fixed, declared set of inputs.

\section{Limitations}

The bootstrap experiment demonstrates a property; it does not resolve all
questions about the technique's generality.

\textbf{Complexity scaling.}
The 926-word specification is simple. Whether the technique scales to
specifications of 10,000 or 100,000 words is an open question. The Attractor
case (34,900 words) provides evidence that longer specifications remain
tractable, but verification difficulty grows: the test suite must cover a
larger behavioral surface, and the specification itself may harbor internal
inconsistencies that only manifest in complex interactions.

\textbf{Model dependency.}
The bootstrap succeeds only with recent state-of-the-art models. Earlier or smaller models
generate either syntactically invalid or behaviorally incorrect implementations.
The bootsrapping technique requires a model capable of faithfully executing a natural-language
specification; this capability is not guaranteed to be stable or universal across model families and has only been reached recently.

\section{Related Work}

The Attractor project by StrongDM~\cite{attractor} applies the same
idea principle at a larger scale. Attractor's specification is
approximately 34,900 words; the specification used in the present experiment
is 926 words, roughly 38 times shorter. The smaller scale makes the present
repository more suitable for learning and for reproducing the bootstrap
property from scratch. A notable feature of the Attractor repository is that
it publishes only the specification, treating the implementation as a
regenerable artifact.
Our accompaniying open-science repository also includes the generated agent code
to serve both as scientific evidence and learning material.

\section{Conclusion}

A coding agent can bootstrap itself. Starting from a natural-language
specification and a first implementation produced by an existing agent, a
newly generated agent re-implements the same specification correctly. The
first self-generated implementation satisfies the same specification as the
first-generation implementation, verified by manual inspection.

This result reproduces, in the domain of AI coding agents, the classical
bootstrap sequence from compiler construction, and instantiates the
meta-circular property from Lisp. The practical implication is a shift in
the focus of engineering effort: the specification is the artifact of record,
and the implementation is regenerable on demand.

\section*{Acknowledgment}

The author thanks StrongDM for the Attractor project, the participants of the
Bellairs 2026 Workshop on Continuous Software Engineering, and the Anthropic AI coding team.
Generative AI has been used to edit this article.
\section*{Author biography}

Martin Monperrus is Professor of Software Technology at KTH Royal Institute
of Technology and IEEE Fellow. 
He has been working on AI coding since his 2009 paper on machine learning for code completion (ACM SigSoft Impact Award 2024). 

\bibliographystyle{plain}
\bibliography{refs}

@article{mccarthy1960,
  author    = {John McCarthy},
  title     = {Recursive functions of symbolic expressions and their computation
               by machine, {Part I}},
  journal   = {Communications of the ACM},
  volume    = {3},
  number    = {4},
  pages     = {184--195},
  year      = {1960},
  doi       = {10.1145/367177.367199}
}

@article{thompson1984,
  author    = {Ken Thompson},
  title     = {Reflections on Trusting Trust},
  journal   = {Communications of the ACM},
  volume    = {27},
  number    = {8},
  pages     = {761--763},
  year      = {1984},
  doi       = {10.1145/358198.358210}
}

@misc{metacircular,
  author       = {Martin Monperrus},
  title        = {The meta-circular coding agent},
  year         = {2026},
  howpublished = {\url{https://github.com/ASSERT-KTH/meta-circular}}
}

@misc{claudecode,
  author       = {{Anthropic}},
  title        = {Claude Code},
  year         = {2025},
  howpublished = {\url{https://claude.ai/code}}
}

@misc{osmani2026,
  author       = {Addy Osmani},
  title        = {Code Review in the Age of {AI}},
  year         = {2026},
  month        = jan,
  howpublished = {Elevate (Substack),
                  \url{https://addyo.substack.com/p/code-review-in-the-age-of-ai}}
}

@misc{attractor,
  author       = {{StrongDM}},
  title        = {Attractor},
  year         = {2026},
  howpublished = {\url{https://github.com/strongdm/attractor}}
}

@misc{lopopolo2026,
  author       = {Ryan Lopopolo},
  title        = {Harness engineering: leveraging {Codex} in an agent-first world},
  year         = {2026},
  month        = feb,
  howpublished = {OpenAI Engineering Blog,
                  \url{https://openai.com/index/harness-engineering/}}
}

@techreport{hartlevin1962,
  author      = {Timothy P. Hart and Mike Levin},
  title       = {The New Compiler},
  institution = {Massachusetts Institute of Technology},
  number      = {AIM-39},
  year        = {1962}
}

@inproceedings{courant2022,
  author    = {Nathana\"{e}l Courant and Justus Decken and
               Simon Castellan and Gabriel Scherer},
  title     = {Debootstrapping without Archeology: Stacked Implementations
               in Camlboot},
  booktitle = {Proceedings of the {ACM} on Programming Languages},
  volume    = {6},
  pages     = {1--29},
  year      = {2022},
  doi       = {10.1145/3498669}
}

@misc{piskala2026,
  author       = {Richard Piskala},
  title        = {Spec-Driven Development: From Code to Contract in the
                  Age of {AI} Coding Assistants},
  year         = {2026},
  howpublished = {arXiv preprint}
}

@article{avizienis1985,
  author  = {Algirdas Avizienis},
  title   = {The {N}-Version Approach to Fault-Tolerant Software},
  journal = {{IEEE} Transactions on Software Engineering},
  volume  = {11},
  number  = {12},
  pages   = {1491--1501},
  year    = {1985},
  doi     = {10.1109/TSE.1985.231893}
}

@article{ron2024,
  author  = {Javier Ron and Diogo Gaspar and Javier Cabrera-Arteaga and
             Benoit Baudry and Martin Monperrus},
  title   = {Gal\'{a}pagos: Automated {N}-Version Programming with {LLMs}},
  journal = {{ACM} Transactions on Software Engineering and Methodology},
  year    = {2024},
  doi     = {10.1145/3785363}
}

@inproceedings{jimenez2024,
  author    = {Carlos E. Jimenez and John Yang and Alexander Wettig and
               Shunyu Yao and Kexin Pei and Ofir Press and Karthik Narasimhan},
  title     = {{SWE-bench}: Can Language Models Resolve Real-World {GitHub} Issues?},
  booktitle = {Proceedings of the International Conference on Learning
               Representations ({ICLR})},
  year      = {2024}
}

\end{document}